\documentclass[aps,prb,twocolumn]{revtex4}
\usepackage{graphicx}

\begin{document}
\draft
\title{Transverse and secondary voltages in
$\rm{Bi_{2}Sr_{2}CaCu_{2}O_{8}}$ single crystals}
\author{K. Vad$^1$, S. M\'esz\'aros$^1$, and B. Sas$^2$}
\address{Institute of Nuclear Research, P.O.Box 51, H--4001 Debrecen,
Hungary\\
$^{2}$ Research Institute of Solid State Physics, P.O.Box 49,
H--1525 Budapest, Hungary}

\date{\today}

\begin{abstract}
Multicontact configuration is one of the most powerful arrangements
for electrical transport measurements applied to study vortex phase
transition and vortex phase dimensionality in strongly anisotropic
high-$T_c$ superconducting materials. In this paper we present electrical
transport measurements using a multiterminal configuration, which prove
both the existence of guided vortex motion in
${\rm{Bi_{2}Sr_{2}CaCu_{2}O_{8}}}$ single crystals near the
transition temperature and that secondary voltage in zero external magnetic
field is induced by thermally activated vortex loop unbinding. The phase
transition between the bound and unbound states of the vortex loops was
found to be below the temperature where the phase coherence of the
superconducting order parameter extends over the whole volume of the
sample. We show experimentally that 3D/2D phase transition in vortex
dimensionality is a length-scale-dependent layer decoupling process and
takes place simultaneously with the 3D/2D phase transition in
superconductivity at the same temperature.
\end{abstract}
\maketitle

\section{Introduction}
In the mixed state of isotropic type II superconductors the vortices can
move in the direction of the Lorentz force due to transport currents. The
angle between the local transport current density and the direction of the
electric field induced by vortex motion is the Hall angle, which can
be determined experimentally by measuring the Hall voltage. If the
superconducting system has structural inhomogeneities, the direction
of vortex velocity can differ from that of the Lorentz force. This
misalignment between the directions of vortex velocity and the Lorentz
force can be identified by observing an even transverse voltage, which, as
opposed to the odd Hall voltage, does not change sign upon magnetic field
reversal. This transverse voltage can be attributed to guided vortex motion,
where vortices move along in a particular direction determined by the
structure of the sample.

Guided vortex motion has been observed in classical type II
superconductors \cite{danckwert} and is expected to be more pronounced
in high-$T_c$ materials. In ${\rm {YBa_{2}Cu_{3} O_{7-\delta}}}$
single crystals it can be induced by twin boundaries, as it has been proved
by magneto-optical technique \cite{vlasko}, magnetization \cite{oussena}
and transverse voltage \cite{chabanenko} measurements. These results
confirm that vortices move in a channel, formed at one edge of a twin
boundary along the twin planes. The other possibility of guided vortex
motion is the intrinsic channeling between CuO$_2$ planes due to intrinsic
pinning. This effect in La-Sr-Cu-O superconductors was observed by
C.A. Dur\'{a}n {\em et al.} \cite{duran} and in
${\rm {YBa_{2} Cu_{3}O_{7-\delta }}}$ films by P. Berghuis
{\em et al.} \cite{berghuis}. Channel formation in 2H-NbSe$_2$ at
relatively low temperatures was experimentally studied by Z. Xiao {\em et
al.} \cite{xiao}. It was shown that depinning of a strongly pinned vortex
lattice starts through the formation of weakly pinned regions in which the
vortices start moving first. In ${\rm {Bi_{2}Sr_{2}CaCu_{2} O_{x}}}$
guided vortex motion was studied by the Hall resistivity measurements in
textured samples \cite{vasek}. To our knowledge, in
${\rm {Bi_{2} Sr_{2}CaCu_{2}O_{8}}}$ single crystals the existence of
guided vortex motion has not been proven yet.

Of high-$T_c$ superconducting compounds
$\rm{Bi_{2}Sr_{2}CaCu_{2}O_{8}}$ is preferred because, due to its
extremely high anisotropy, its discrete superconducting layers play an
important role in current conduction properties even in the transition
temperature range, and, due to the highly anisotropic conductivity, the
vortex matter has a very rich phase diagram with numerous phase
transitions. The multicontact configuration is one of the most promising
arrangements of electrical transport measurements in the study of
superconducting phase transition and vortex dimensionality. In this
configuration four electrical contacts are attached to one side of a single
crystal and two or four contacts to the opposite side. If the current injected
into one side of a crystal is high enough, a voltage drop on both sides can
be measured. The side where the current is injected is called primary, the
opposite side is secondary. Due to this contact arrangement and the
anisotropic conductivity, a non-uniform current distribution develops in the
sample, and the current injected into the primary side is confined to a very
thin surface layer, which is thinner than the thickness of the crystal. If the
pancakes in adjacent layers belonging to the same vortex line are strongly
coupled to each other, the primary current induces vortex motion not only
in the primary, but also in the secondary surface, causing non-local
dissipation in the sample. Moreover, if the coupling strength between the
pancakes is high enough to produce three-dimensional (3D) type vortex
lines across the sample, primary and secondary voltages can be equal. This
configuration reminds us of the geometry used by I. Giaever in his famous
experiment \cite{giaever}, which is called dc flux transformer
configuration.

The first measurements on $\rm{Bi_{2}Sr_{2}CaCu_{2}O_{8}}$ single
crystals using the dc flux transformer configuration were performed by H.
Safar {\em et al}. \cite{safar} and R. Busch {\em et al}. \cite{busch}.
They observed the dimensionality of the vortex system over a wide range
of the phase diagram. Y.M. Wan {\em et al}. \cite{wan1,wan2} and C.D.
Keener {\em et al}. \cite{keener} also used the multicontact dc flux
transformer configuration to study the secondary voltage and found that in
magnetic fields near the transition temperature the interlayer vortex coupling
was responsible for the secondary voltage. However, the origin of the
secondary voltage in zero external magnetic field still remained problematic.
S.W. Pierson \cite{pierson1} suggested that it originated from thermally
activated vortex loop unbinding. A vortex loop is a correlated
vortex-antivortex line pair. Using a real-space renormalization group
analysis, the author identified three characteristic critical currents and
calculated their temperature dependence. He found that the temperature
dependence of the secondary voltage is a horizontal slice of the
current-temperature phase diagram.

In this paper we present electrical transport measurements using a
multiterminal configuration that prove the existence of guided vortex
motion in chemically and mechanically homogeneous
${\rm{Bi_{2}Sr_{2}CaCu_{2}O_{8}}}$ single crystals. We also
present the implications of experimental results for the 3D/2D phase
transition in vortex dimensionality near the Ginzburg-Landau transition
temperature and show that this phase transition is a
length-scale-dependent layer decoupling process. We show that the
temperature dependence of the secondary voltage in zero magnetic field
has a double peak structure.

\section{Experimental arrangement}
Single-crystalline $\rm{Bi_{2}Sr_{2}CaCu_{2}O_{8}}$ compounds
were prepared by the melt cooling technique, described in
Reference \cite{keszei}. Optically smooth rectangular crystals were
carefully cleaved from these compounds, and heated in flowing oxygen for
15 minutes at 900 K in order to stabilize the oxygen content. Chemical
homogeneity of the samples was checked by microbeam PIXE and
(O$^{16}, \alpha $) resonant elastic scattering \cite{simon} with a spatial
resolution of 5 $\mu $m. No chemical inhomogeneities were identified. The
surface smoothness was measured by atomic force microscopy. The
surfaces were found to be flat with a typical roughness of 10 nm. Before
preparing the electrical contacts, the sample quality was checked by
magnetization measurements using a SQUID or a vibrating sample
magnetometer, and by AC susceptibility measurements. Electrical contacts
were made by bonding 25 $\mu $m gold wires with Dupont 6838 silver
epoxy fired for five minutes at 900 K. The contact resistance was a few
ohms. The geometrical position of electrical contacts was precisely
measured by optical microscope.

\begin{figure}
\includegraphics[width=6cm, height=3.5cm]{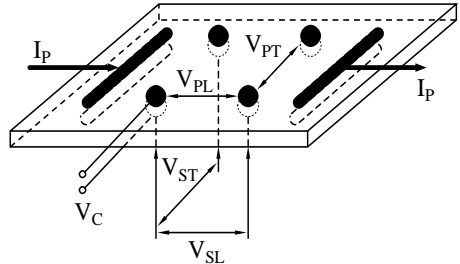}
\caption{Electrode configuration used for transverse and secondary
voltage measurements. For definition of voltages see text. $I_P$
denotes the primary current.}
\label{fig1}
\end{figure}

Two current and four potential electrodes were attached to both faces of
the crystals. The scheme of the electrode configuration is shown in
Fig.\ \ref{fig1}. The current was injected into one face of the single crystal
through the current contacts (this is the primary current $I_P$), while
primary and secondary longitudinal voltages, primary and secondary
transverse voltages, and the voltage between the two faces were recorded
simultaneously, using a six-channel data acquisition system. Applying this
symmetrical contact arrangement, this configuration made it possible to
check the sample homogeneity from the point of view of electrical
conductivity.

We have performed measurements on a number of crystals, but in this
paper we present the results we measured on two of them fabricated from
the same batch (samples $A$ and $B$). The mean-field Ginzburg-Landau
(GL) transition temperatures $T_{c0}$ of the samples were 88.5 K and
88 K, respectively. The sample dimensions were about $1\times 1.5$ mm
$^2$, the thickness was 8 and 3 $\mu $m, respectively. As we pointed out
in our previous papers \cite{sas}, most of the Joule heat due to the
transport current was generated in the current contacts. In order to reduce
the heat dissipation, and also to eliminate the thermoelectric force, we used
current pulses with a duration time of 1 ms and repetition time of 100 ms.
This arrangement and the fact that the sample temperature was regulated by
a temperature controlled He gas stream instead of exchange gas, made it
possible to avoid heating during the pulse up to the amplitude of 10 mA.

In our experiments the following voltages, shown in Fig.\ \ref{fig1}, were
simultaneously recorded: (i) longitudinal voltages measured parallel to the
current direction on the surface of the crystal where the current was
injected and on the opposite surface ($V_{PL}$ and $V_{SL}$, the
primary and secondary longitudinal voltages, respectively); (ii) the primary
and secondary transverse voltages, measured perpendicularly to the current
direction ($V_{PT}$ and $V_{ST}$); (iii) the $c$-axis direction voltage
measured between the two surfaces of the crystal ($V_{C}$).

Using the analysis of R. Busch {\em et al}. \cite{busch}, we could define
the temperature dependence of the $ab$ plane and $c$-axis resistivities
from the voltages $V_{PL}$ and $V_{SL}$. For the anisotropy ratio
$\gamma$ we received
$\gamma =\sqrt(\rho _{c}/\rho _{ab}) \approx 500$ with
$\rho _{ab} \approx 100$ $\mu \Omega$cm at 90 K.

\section{Experimental results}
\begin{figure}
\includegraphics[width=6.5cm, height=8cm]{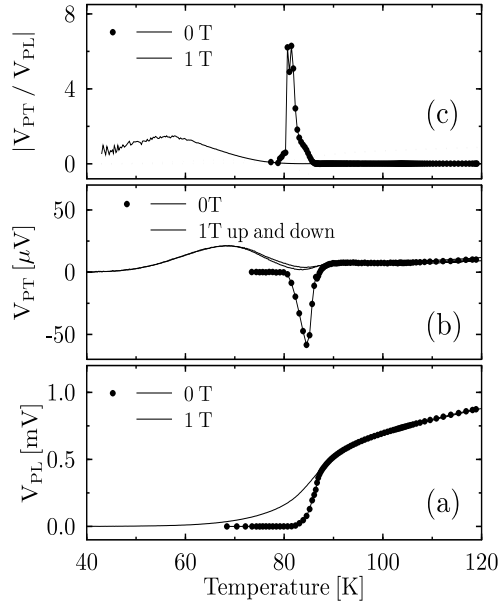}
\caption{Temperature dependence of the primary longitudinal voltage
$V_{PL}$ (a), primary transverse voltage $V_{PT}$ (b) and the absolute
value of the $V_{PT}/V_{PL}$ ratio (c) of a
$Bi_{2}Sr_{2}CaCu_{2}O_{8}$ single crystal measured in zero and 1 T
magnetic fields. In (b) $V_{PT}$ is measured in two opposite magnetic
field directions. The magnetic field is perpendicular to the $ab$ planes,
$I_P$= 1 mA.}
\label{fig2}
\end{figure}

The temperature dependence of the primary longitudinal and transverse
voltages, $V_{PL}$ and $V_{PT}$, of a
${\rm{Bi_{2}Sr_{2} CaCu_{2}O_{8}}}$ single crystal (sample $A$)
measured in zero and 1 T magnetic fields is shown in Fig.\ \ref{fig2}(a and
b). In zero magnetic field $V_{PT}$ shows a sharp change in the transition
temperature range with a sign reversal. This sign reversal also exists in
weak magnetic fields and seems to be a universal characteristic of vortex
motion. Fig.\ \ref{fig2}(c) shows the calculated ratio of the measured
primary transverse and longitudinal voltages in zero and 1 T magnetic
fields. The maximum value of this ratio in zero magnetic field is 6, and it is
larger than 1 in a temperature range of $\sim 3$ K. In 1 T magnetic field
the maximum value is only 1.5 and the temperature range where the ratio is
larger than 1 is 14 K wide. The Hall voltage can be calculated from the
difference of the two transverse voltages measured in a magnetic field
(here 1 T) with opposite field directions (Fig.\ \ref{fig2}(b)).

\begin{figure}
\includegraphics[width=6.5cm, height=7cm]{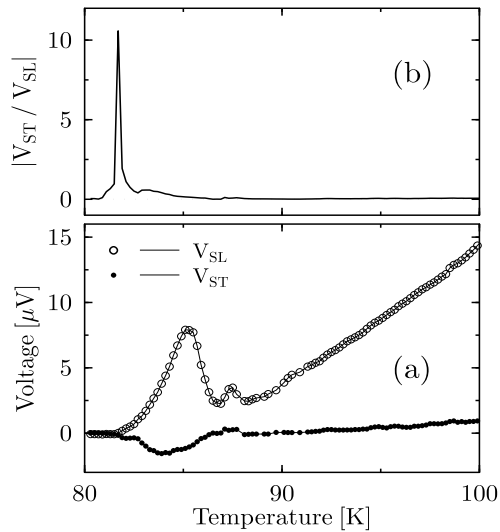}
\caption{Temperature dependence of the secondary longitudinal
$V_{SL}$ and transverse $V_{ST}$ voltages (a) and the absolute value
of the $V_{ST}/V_{SL}$ ratio (b), $I_P$=1 mA.}
\label{fig3}
\end{figure}

\begin{figure}
\includegraphics[width=6cm, height=4.5cm]{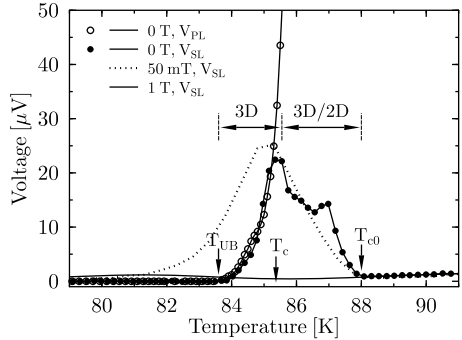}
\caption{Secondary and primary voltages vs. temperature in different
magnetic fields (sample $B$). The magnetic field is parallel to the $ab$
planes, $I_{P}$ = 1 mA. $T_{UB}$, $T_c$ and $T_{c0}$ denote the
unbinding, transition and Ginzburg-Landau transition temperature,
respectively. The different vortex dimensionalities are also shown.}
\label{fig4}
\end{figure}

On the secondary side of the crystal we found that in zero magnetic field
the ratio of transverse and longitudinal voltages, $V_{ST}$ and
$V_{SL}$, is higher than 1 in a temperature range of $\sim 0.7$ K
(Fig.\ \ref{fig3}). This range is about 4 times narrower than it is on the
primary side. In higher magnetic fields, with the field perpendicular to the
$ab$ plane, we could not determine this ratio because both $V_{ST}$ and
$V_{SL}$ decreased to lower than 100 nV at 5 mA measuring current
already in 100 mT magnetic field. Sample $B$, prepared to study the
parallel case, was placed  in the cryostat so that the $ab$ planes were
parallel to the magnetic field direction. In this arrangement $V_{SL}$ is
detectable in high magnetic fields. The result is presented in Fig.\
\ref{fig4}.

\begin{figure}
\includegraphics[width=6cm, height=9cm]{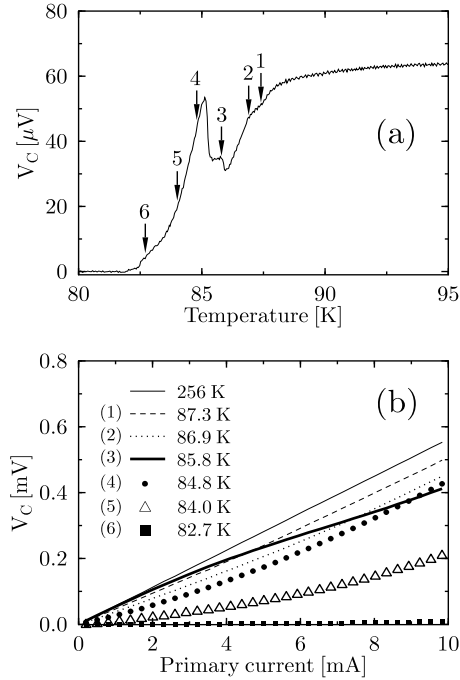}
\caption{Temperature dependence of $V_{C}$ at 1 mA primary current
(a), and current-voltage characteristics measured at different temperatures
(b) in zero magnetic field. The arrows in (a) denote the temperatures where
the current-voltage characteristics were measured.}
\label{fig5}
\end{figure}

In zero applied magnetic field the secondary voltage is determined by
interlayer vortex coupling. In order to gain information about this coupling
strength, we measured the voltage $V_C$ between the two surfaces of the
crystal while applying a primary current $I_P$. Depending on the current
distribution in the crystal, some part of this current flows in the $c$-axis
direction and $V_C$ depends on this current. The temperature
dependence of $V_C$ is of metallic ($d\rho/dT>0$) type, except in a small
temperature range near the mean-field Ginzburg-Landau transition
temperature. In Fig.\ \ref{fig5}(a) this range is between 85 and 86 K. This
temperature dependence is also reflected in the current-voltage
characteristics (Fig.\ \ref {fig5}(b)). Far above the GL transition
temperature, at 256 K the current-voltage characteristic is linear. Near the
GL transition temperature, at 87.3 K there is a slight curvature in it. On
further cooling the sample from 87.3 K to 86 K the curvature increases. In
the temperature range between 86 K and 85 K the $I_{P}-V_C$ curves are
concave in shape (for clarity in Fig.\ \ref{fig5} we present only the curve
measured at 85.8 K). At temperatures lower than 85 K both the
$V_{C}-I_{P}$ and $V_{C}-T$ curves show the same metallic
behaviour, as above 86 K.

\section{Discussion}
\subsection{Guided vortex motion}
The experimental results discussed above can help us to understand the
origin of transverse voltages. In general, transverse voltage consists of
three components: a component coming from the geometrical misalignment
of the contacts, a Hall voltage component, and a component originating
from the guided motion of vortices. The geometrical misalignment can be
detected by normal-state measurements. The other two components can be
distinguished by comparing the voltages measured in two opposite
magnetic field directions: while in the Hall voltage the sign of the measured
voltage depends on the external magnetic field direction because the vortex
motion is only governed by the Lorentz force, in the case of the guided
vortex motion it is independent of the field direction. This difference gives
us the chance to distinguish between these two voltage generation
mechanisms. On the basis of our results we can conclude that the guided
motion of vortices is reflected in both $V_{PT}$ and $V_{ST}$
transverse voltages (Fig.\ \ref{fig2} and \ref{fig3}), and that in our samples
the guided vortex motion is a dominating effect in the Hall voltage
measurements, since the Hall voltage is superimposed on a voltage, which
is found to be much higher than the Hall voltage itself (Fig.\ \ref{fig2}(b)).

In our electrode configuration the main part of the injected current $I_P$
flows in the $ab$ planes, but has a component in the $c$-axis direction,
too. If the external magnetic field is parallel to the $c$-axis, the latter
component of the current is parallel to the vortex lines, so, it does not
contribute to the Lorentz force. The dominating component of the
transverse voltage is brought about by the intralayer current.

There is a well-defined temperature range, where the ratio of transverse and
longitudinal voltages is larger than 1, while the geometrical misalignment
between the transverse voltage contacts is less than 10 \% of the distance
between the longitudinal voltage contacts, as it was measured by an optical
microscope. This supports the view that the transverse voltage cannot
originate from a geometrical misalignment of the potential contacts.
Chemical inhomogeneity does not explain this phenomenon, either. We
believe that these results can be understood within the concept of guided
vortex motion. If the magnetic field is perpendicular to the $ab$ planes,
slight inhomogeneities in pinning potentials can bring about channels for
moving vortices, which results in guided motion. Some vortices are
localized by stronger pinning centers and form trapped vortex regions,
other mobile vortices flow in channels between these trapped regions. The
existence of such vortex channels decreases the critical current because the
effective pinning potential is lower, consequently, vortex mobility is higher
in a channel than outside. Guided vortex motion is sensitive to applied
magnetic fields. Higher magnetic fields restricts the guided motion of
vortices through vortex-vortex interaction.

\subsection{Secondary longitudinal voltage}
In 2D superconducting layers the phase fluctuation of the order parameter
generates vortex-antivortex pairs as topological excitations\cite{tinkham}.
The phase transition in an isolated 2D superconducting layer, where the
vortex-antivortex pairs are bound below the phase transition temperature
and are unbound above it, is described by the Kosterlitz-Thouless theory
\cite {kosterlitz,halperin}. In thin films, if the film thickness is less than the
superconducting coherence length, this Kosterlitz-Thouless phase
transition ($T_{KT}$) can be observed.

In high-temperature superconductors, the high transition temperature, short
coherence length and layered structure make the phase fluctuation of the
order parameter dominant over the other fluctuations in the transition
temperature range, but the coupling between superconducting layers
modifies the 2D Kosterlitz-Thouless picture. The interaction between
superconducting layers leads to vortex-antivortex interaction different from
the 2D case. In $\rm{Bi_{2}Sr_{2}CaCu_{2}O_{8}}$ single crystals the
CuO$_2$ planes serve as 2D superconductor layers, the vortices are the
pancake vortices. Due to the layered structure a high anisotropy exists in
conductivity and the vortex matter has a very rich phase diagram with
numerous phase transitions. Among others the 3D phase appears, where
the coupling between neighbouring layers arranges the thermally excited
pancake vortices into 3D flux lines\cite{blatter}. These 3D flux lines can
form vortex loops as correlated vortex-antivortex line pairs. They are called
thermally activated vortex loops, because they are the result of a combined
effect of thermally activated vortex excitation and interlayer vortex
coupling.

The 3D character modifies the structure of the phase transition
between the bound and unbound states. Although the 2D signatures are
prevalent, a narrow 3D window appears around the phase transition
temperature ($T_c$) and a nonzero critical current appears. The 3D
temperature region is theoretically predicted and it has been shown that the
size of this 3D region is much smaller above $T_c$ than
below\cite{pierson2}. Above the phase transition temperature the
behaviour of vortices becomes 2D due to decoupling of the
superconducting layers (3D/2D phase transition). It has been also shown in
another paper\cite{pierson3} that this layer decoupling is a
length-scale-dependent process: the layers become decoupled at length
scales larger than an interlayer screening length, while for lengths below this
scale they remain coupled. The manifestation of the 3D/2D transition in
electrical transport properties has not been investigated before.

In an isolated 2D superconducting layer the vortex pair energy is the
sum of the creation energy of the two vortices and the intralayer logarithmic
coupling energy between vortices. In layered systems like
$\rm{Bi_{2}Sr_{2}CaCu_{2}O_{8}}$ this vortex pair energy is modified
by the interlayer Josephson coupling, which not only strengthens the
intralayer interaction, but causes an interaction between vortices located in
neighbouring layers. The intralayer vortex pair energy can be written in the
following simple form:
$E(r)=2E_{c}+K_{\parallel} ln(r/\xi_{0})+K_{\perp} r/\xi_{0}$,
where $r$ is the distance between the vortex and antivortex, $E_{c}$ is the
creation energy of a vortex, $\xi_{0}$ is the zero-temperature correlation
length, $K_{\parallel}$ is the intralayer vortex-vortex coupling constant and
$K_{\perp}$ is the interlayer Josephson coupling constant. While the 2D
logarithmic interaction (second term in this equation) dominates at short
distances, the Josephson-coupling mediated 3D linear interaction (third
term) dominates at large distances. The intralayer vortex length scale
$R_{\lambda}(T)$ is the characteristic length which divides this
logarithmic and linear regimes\cite{pierson3}.
$R_{\lambda}(T)=\xi_{0}/\sqrt \lambda$, where $\lambda$ is the ratio of
the interlayer Josephson coupling to the intralayer coupling,
$\lambda=K_{\perp}/K_{\parallel}$. $\lambda$ depends on the size $r$ of
the vortex pairs, and above $T_c$ it has a maximum. The size $r$ which
belongs to this maximum $\lambda$ is the other characteristic length, the
interlayer screening length ${\it l}\,_{3D/2D}(T)$. If the separation
between two vortices located in neighbouring layers is larger than
${\it l}\,_{3D/2D}$, the Josephson-coupling mediated linear interaction is
screened out and the layers are decoupled. Around $T_c$ the dimensional
behaviour of the system is determined by these two characteristic lengths.
The layers are coupled and the behaviour of vortices is 3D if the separation
between vortices is greater than the intralayer vortex length scale
$r>R_{\lambda}(T)$ and less than the interlayer screening length
$r<{\it l}\,_{3D/2D}(T)$. The temperature dependence of
$R_{\lambda}(T)$ and ${\it l}\,_{3D/2D}(T)$ shows\cite{pierson4} that
the intralayer vortex length scale $R_{\lambda}(T)$ is constant for
$T \ll T_c$, but it increases as the temperature approaches $T_c$ from
below. The interlayer screening length ${\it l}\,_{3D/2D}(T)$ decreases
continuously as the temperature increases.

While in the 3D regime below $T_{c}$ the electrical transport behaviour is
dominated by vortex loops, above $T_{c}$ it is dominated by vortex lines
and pancake vortices. The multiterminal configuration is a good
arrangement to distinguish between these two regimes. In this paper
we show that studying the temperature dependence of the secondary
voltage can help us to understand the length-scale-dependent layer
decoupling, i.e., the 3D/2D phase transition.

The secondary voltage has already been studied both theoretically
\cite{pierson1} and experimentally\cite{wan1}. It can now be accepted that
in zero applied magnetic field it originates from thermally activated vortex
loop unbinding. At low temperatures where $V_{SL}$ and $V_{PL}$ are
zero, the thermally excited 3D flux lines form vortex loops which are
'pinned' to the crystal. With increasing temperature, the transport current
splits these vortex loops into free vortex-antivortex line pairs. The
temperature where this splitting starts is the unbinding temperature
($T_{UB}$). This is the lowest temperature where both $V_{SL}$ and
$V_{PL}$ are observable. Above the unbinding temperature the free
vortices move in the sample like 3D vortex lines due to the Lorentz force,
producing the same voltage drop on the primary and secondary side of the
crystal, $V_{PL}=V_{SL}$. This 3D character of the vortex lines remains
up to a temperature where the secondary voltage has a local maximum. At
sample $B$ the 3D temperature range is around 85 K where the zero field
primary and secondary longitudinal voltages $V_{PL}(0T)$ and
$V_{SL}(0T)$ coincide (see Fig.\ \ref{fig4}). This is the same 3D
temperature range which was predicted theoretically by renormalization
group analysis \cite{pierson2}. With increasing temperature the 3D
character of flux motion disappears and consequently $V_{SL}$ becomes
lower than $V_{PL}$, but another local maximum of $V_{SL}$ can be
found as the temperature approaches $T_{c0}$. The temperature
dependence of the secondary voltage has two peaks with a higher and a
lower amplitude.

This double peak structure of $V_{SL}(0T)$ can be explained by the
motion of different types of vortex lines. In zero applied magnetic field free
vortex lines can be produced in two ways. First, they can be the result
of vortex-antivortex depairing of thermally activated vortex loops due
to the Lorentz force of the transport current. In this case the number of
free vortex lines depends on the transport current density and a non-Ohmic
behaviour characterizes the system. Secondly, free vortex lines can be
spontaneously created by thermal activation, mainly above $T_{c}$. The
number of free vortex lines increases with increasing temperature and the
system is characterized by an Ohmic behaviour.

The effect of current on vortex-antivortex depairing is twofold. On the one
hand the current reduces the creation energy whereby increases the density
of vortex pairs. On the other hand the current exerts a force (the Lorentz
force) on vortex loops and can blow them out. The number of blowouts
depends on the size $r$ of the vortex pairs. If $r$ is higher than the
intralayer vortex length scale $R_{\lambda}(T)$, the Josephson-coupling
mediated 3D linear interaction is energetically favoured over 2D intralayer
logarithmic interaction and the energy of a vortex loop is smaller than the
energy of a pair of independent vortex lines. Below $T_c$ in
$\rm{Bi_{2}Sr_{2}CaCu_{2}O_{8}}$
$R_{\lambda}(T)=\xi_{0}/\sqrt \lambda \approx 1 \mu$m, where
$\xi_{0} \sim 3$nm and $\lambda \sim 10^{-5}$. Consequently, if
$r>1 \mu$m, creation of vortex loops is energetically favoured over free
vortex-antivortex line pairs. This happens in the 3D temperature range
where the dominant topological excitation is the vortex loop. With
increasing temperature the number of blowouts and, so, the longitudinal
secondary voltage increases. However, approaching the transition
temperature the intralayer vortex length scale $R_{\lambda}(T)$ starts to
increase, therefore the number of vortex pairs which can be blown out by
the transport current decreases which results in the decrease of the
secondary voltage. The temperature which belongs to the peak value of the
secondary voltage is the transition temperature $T_c$. Although at this
temperature the behaviour of vortices is still 3D type as it was shown
theoretically\cite{pierson4}, the secondary voltage is somewhat lower than
the primary voltage (see Fig.\ \ref{fig4}). This means that the secondary
voltage starts to decrease before layer decoupling. At higher temperatures
the 3D character of flux motion disappears and $V_{SL}$ becomes
significantly lower than $V_{PL}$. Above the 3D temperature range
another local maximum of $V_{SL}$ can be found as the temperature
approaches $T_{c0}$, because the temperature is high enough to
produce free vortex lines by thermal activation. The number of free vortex
lines and, so, the secondary voltage increases with increasing temperature.
The system is characterized by a continuous 3D/2D transition due to
continuous decrease of the interlayer screening length
${\it l}\,_{3D/2D}(T)$ as the temperature approaches the mean-field
Ginzburg-Landau transition temperature $T_{c0}$. Near $T_{c0}$ the
amplitude of the order parameter decreases, just as the number of the free
vortex lines. This effect evokes the decrease of the secondary voltage. The
temperature which belongs to the minimum secondary voltage above the
double peak structure is $T_{c0}$.

The local maxima of $V_{SL}$ can be decreased by magnetic field. If the
magnetic field is perpendicular to the $ab$ planes, the external field
prevents the free motion of vortex-antivortex line pairs. While one part of a
pair, which is parallel to the external magnetic field, cannot move in the
sample because of the vortex-vortex interaction, the other antiparallel part
is annihilated in the external magnetic field. If the magnetic field is parallel
to the $ab$ planes, there is no interaction between the vortex lattices
formed by the external magnetic field and vortex loops, and the secondary
voltage is not so sensitive to the applied magnetic field. In this arrangement
the magnetic field reduces the coupling strength between the CuO$_2$
bilayers by decreasing both the magnetic and Josephson coupling. This has
a twofold consequence. First, the thermally excited 2D vortex-antivortex
pair creation in CuO$_2$ bilayers is suppressed by magnetic field because
the creation of a vortex-antivortex pair is equivalent to the creation of a
dislocation in the interlayer flux line lattice which increases the elastic
energy of the lattice. Secondly, the magnetic field reduces the probability
of arrangement of thermally excited pancake vortices into 3D flux lines.
That is why only one peak exists in low magnetic fields (50 mT) instead of
a double peak (Fig.\ \ref{fig4}).

Information about the strength of Josephson coupling and the 2D/3D phase
transition in superconductivity can be obtained by study the temperature
dependence of $I_{P}-V_C$. In the temperature range where the
$I_{P}-V_C$ curves are concave (between 85 and 86 K) the
superconducting coherent state exists in the CuO$_2$ bilayers, but the
interlayer coupling is not strong enough to establish the $c$-axis
coherence. In consequence, the current in $c$-axis direction is carried by
single particle tunnelling instead of Cooper pair tunnelling (2D type
superconductivity). With the decrease in temperature, at 85 K, the
Josephson coupling, and consequently, a phase coherence between
superconducting bilayers develops and extends over the whole volume of
the sample, producing maximum values in both $V_{SL}$ and $V_C$
(2D/3D phase transition). At temperatures lower than 85 K the Josephson
coupling energy increases, the phase coherence of the superconducting
order parameter extends over the whole volume of the sample and
develops the 3D type superconductivity. The higher peak value of
$V_{SL}$ is also at 85 K which is the upper end of the temperature range
where vortex lines have 3D character. Consequently 2D/3D phase transition
in superconductivity and in vortex dimensionality takes place at the same
temperature. This can be valid inversely, too. If the vortex dimensionality
decreases from 3D to 2D, the dimensionality of superconductivity can also
decrease. This result was experimentally supported by transport current
measurements\cite{pethes}, where the authors proved that
the Bardeen-Stephen model for the flux flow resistance
$\rho _{f}=\rho _{n}\cdot B/B_{c2}$ ($\rho _{n}$, $B$ and $B_{c2}$
are the normal state resistivity, magnetic field and critical magnetic field) is
not valid at high current density in $\rm{Bi_{2}Sr_{2}CaCu_{2}O_{8}}$
single crystals. Due to intensive flux motion both the phase coherence
between superconducting bilayers and the interlayer screening length
${\it l}\,_{3D/2D}(T)$ decrease resulting in a 3D/2D phase transition in
dimensionality of superconductivity.

Thermally unbound vortex-antivortex line pairs take part in guided motion.
The maximum values of $\mid V_{PT}/V_{PL}\mid $ and
$\mid V_{ST}/V_{SL}\mid$ are near the unbinding temperature in the 3D
type superconducting regime. So, the guided vortex motion has a 3D
character in $\rm{Bi_{2}Sr_{2}CaCu_{2}O_{8}}$ single crystals and
develops due to melting of vortex loop lattice. The external magnetic field
which is perpendicular to the $ab$ planes restricts the guided motion of
vortices through vortex-vortex interaction. That is why weak magnetic
fields can prevent the development of secondary transverse voltage.

\section{Conclusions}
In conclusions, we found that temperature dependence of secondary
longitudinal voltage has a double-peak structure and its higher maximum
value is at the temperature where the phase coherence of the order
parameter extends over the whole sample. Secondary
voltage originates from correlated vortex-antivortex line pair unbinding, i.e.,
from vortex loop unbinding due to the Lorentz force of the transport
current. Near $T_{c0}$ free vortex-antivortex line pairs are also generated
by thermally activated vortex excitation. We think that the two types of
vortex-antivortex line pairs are responsible for the double peak structure of
the secondary longitudinal voltage. Lacking of theories describing the
double peak structure in $\rm{Bi_{2}Sr_{2}CaCu_{2}O_{8}}$ single
crystals indicates a need for better description of length-scale dependence
in layered superconductors. In order to improve the theoretical description,
one can perform the renormalization group analysis of Reference
\cite{pierson5} for non-constant current density or one can use different
renormalization group methods, e.g., field theoretical renormalization group
approaches\cite{nandori}.

\section{Acknowledgement}
We take great pleasure in acknowledging discussion with P.F. de
Ch\^{a}tel and I. N\'{a}ndori. This work was supported by the Hungarian
Science Foundation (OTKA) under contract no. T037976.

\end{document}